\documentclass[final,3p,times]{elsarticle}

\usepackage{amssymb}
\usepackage{amsmath}
\usepackage{amssymb}
\usepackage{latexsym}
\usepackage{epsfig}
\usepackage{amsbsy}

    \usepackage{subfig}
    \usepackage{booktabs}
    \usepackage{epsfig}
    \usepackage[justification=RaggedRight]{caption}
    \usepackage{hyphenat}
    \usepackage{multicol}

\journal{Physica A}

\begin{document}

\begin{frontmatter}



\title{Quantum spatial-periodic harmonic model for daily price-limited stock markets}


\author[label1]{Xiangyi Meng}
\author[label2]{Jian-Wei Zhang}
\author[label3]{Jingjing Xu}
\author[label1]{Hong Guo\corref{CorrespondingAuthor}}
\cortext[CorrespondingAuthor]{Tel:+86-10-6275-7035}
\address[label1]{State Key Laboratory of Advanced Optical Communication$\text{,}$ Systems and Networks$\text{,}$ School of Electronics Engineering and Computer Science$\text{,}$ and Center for Quantum Information Technology, Peking University, Beijing 100871, China}
\address[label2]{School of Physics, Peking University, Beijing 100871, China}
\address[label3]{School of Mathematical Sciences, Peking University, Beijing 100871, China}
\ead{hongguo@pku.edu.cn}
\date{}
\begin{abstract}
We investigate the behavior of stocks in daily price-limited stock markets by purposing a quantum spatial-periodic harmonic model. The stock price is presumed to oscillate and damp in a quantum spatial-periodic harmonic oscillator potential well. Complicated non-linear relations including inter-band positive correlation and intra-band negative correlation between the volatility and the trading volume of stocks are derived by considering the energy band structure of the model. The validity of price limitation is then examined and abnormal phenomena of a price-limited stock market (Shanghai Stock Exchange) of China are studied by applying our quantum model.
\end{abstract}

\begin{keyword}
Econophysics \sep Quantum harmonic model \sep Price-limited stock market \sep Volatility \sep Trading volume \sep Energy band structure


\PACS 89.65.Gh \sep 05.40.Jc \sep 03.65.Yz

\end{keyword}

\end{frontmatter}

% \linenumbers

\section{Introduction}
\label{Section1}
Stock market, as one of the most important financial instruments, plays an unshakable role in basic research of finance and economics. After the noticeable works of relating economic research with fundamental concepts and methods of statistic physics in the 90s \cite{Stanley2,Stanley}, econophysics soon burgeons as a new interdisciplinary area, from which quantum finance is then specifically introduced for applying quantum physics to finance \cite{QuantumFinancePathIntegrals, PRE2,
QuantumFinance, PersistentFluctuationsStockMarkets, QuantumStatisticalStockMarkets, FuDanQuantumModel, MinimalLengthUncertainty, QF1, QF2, QF3, QF4, QBM}. With help of quantum mechanics \cite{QuantumMechanicsCT}, it is remarked that a single stock can be treated as a quantum harmonic oscillator, excited by external information, meanwhile damping to its ground state \cite{PersistentFluctuationsStockMarkets}, while the stock index behaves as a quantum Brownian particle with assemble of stocks as a thermal reservoir \cite{QBM}, of which a quantum Brownian model is introduced in order to explain fat tail phenomena \cite{ZJW4} and long-term non-Markovian features \cite{ZJW1} by applying the theory of quantum open systems \cite{OpenQuantumSystems}. By combining physics models with financial tools, we are able to study the underlying physical concepts of finance and economics and handle with financial problems more effectively.

It is worth noticing that the stock markets of China--an economically burgeoning developing country--are much more concerned and being studied in recent years \cite{StockMarketChina1, StockMarketChina2, StockMarketChina3}. One of the particularities of the stock markets in China is daily price limitation, i.e., the daily increase (decrease) of one stock price is limited in $\pm 10\%$ ($\pm 5\%$ for special treated stocks) \cite{PriceLimit1, PriceLimit2, PriceLimit3, PriceLimit4, PriceLimit5}. After reaching the limit, transaction of the stock is not paused, but is limited in one direction, and is thus strongly biased to sellers (buyers). We also notice a recent incident of the stock markets in China, that on Aug. 16, 2013, a so-called fat finger transaction incident caused by China Everbright Securities lead to unusual fluctuations of the stock markets. The stock index rose $+5.62\%$ with more than thirty numbers of stocks reaching their $+ 10\%$ price limits rapidly, and then fell down as soon as possible in just one minute. The volatility is so large that it is considered to be statistically abnormal. Also, the incident is related with price limitation, and such an eccentric phenomenon requires a more sophisticating investigation. Since the influence of price limitation on stock markets is still questioned and argued, we remark that price limitation remains to be a valuable issue for research, which will help us to reconsider the applicability of price limitation and predict its further influence on stock markets.

In this paper, we investigate the price-limited space of stock price with the help of quantum energy band theory and introduce a spatial-periodic harmonic oscillator potential well in the space. The theoretical model is derived which implies the existing of an energy band structure in price-limited stock markets, and the energy band structure will introduce complicated relations between volatility and trading volume. From a more detailed numerical solution of the spatial-periodic harmonic model, the exact solution of a non-linear relation between volatility and trading volume is derived, which implies that not only ordinary inter-band positive correlations but also abnormal intra-band negative correlations are contained in the complex relation. The ability for price limitation to limit volatility is then reconsidered, which implies that the price limit will increase the volatility if within a certain regime of the trading volume. Hence, abnormal phenomena and features of price-limited stock markets are able to be explained and predicted, which will provide better regulatory methods for stock markets. We remark that the physic model introduced by us sets up a new attracting point of view, and contributes to the development of econophysics as well as quantum finance.

\section{Quantum spatial-periodic harmonic oscillator potential well}
\label{Section2}
\begin{figure}[!thbp]
\begin{centering}
\begin{multicols}{2}
\subfloat[(b)]{\begin{centering}
\includegraphics[width=8.0cm]{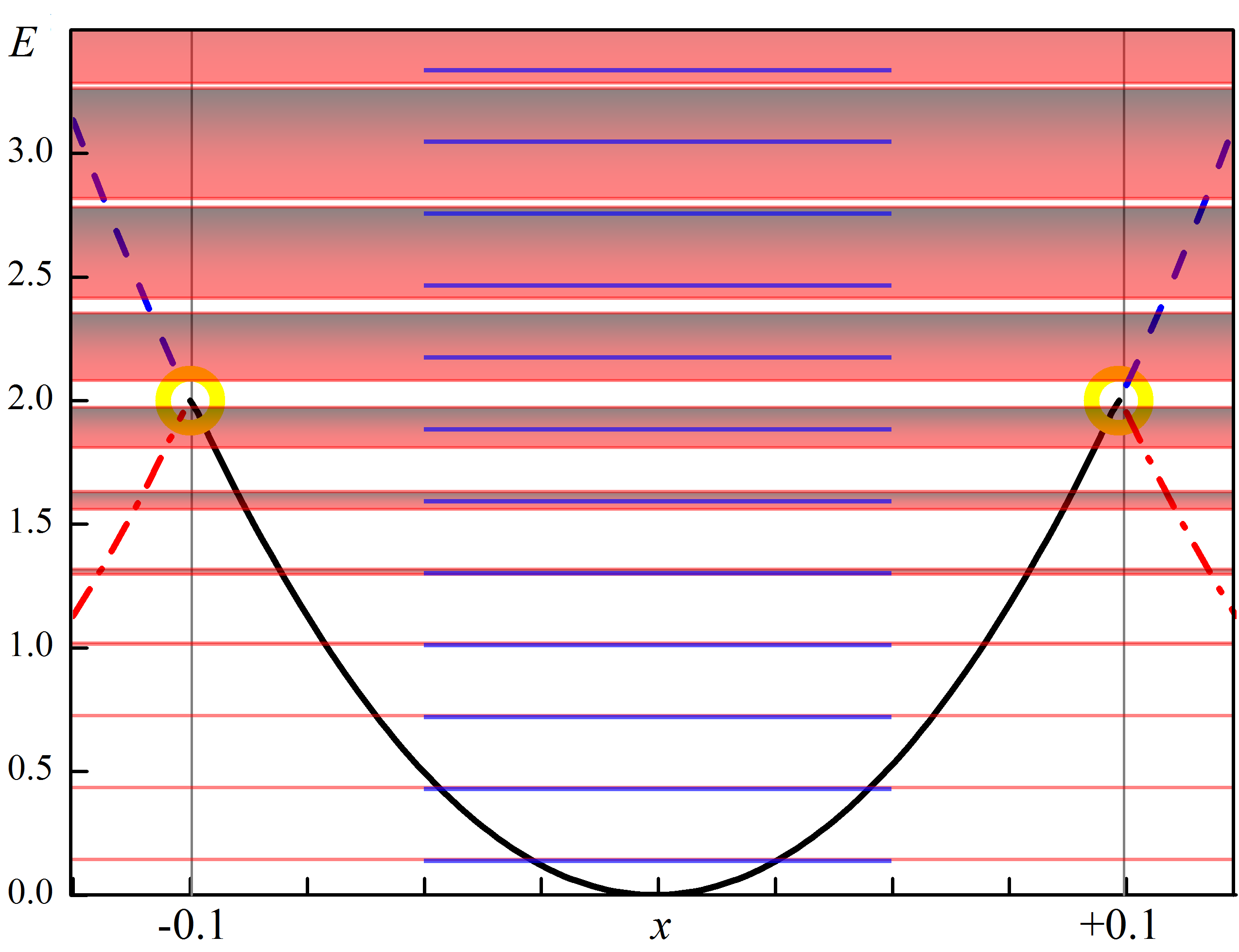}
\par\end{centering}
}

\subfloat[(c)]{\begin{centering}
\includegraphics[width=8.2cm]{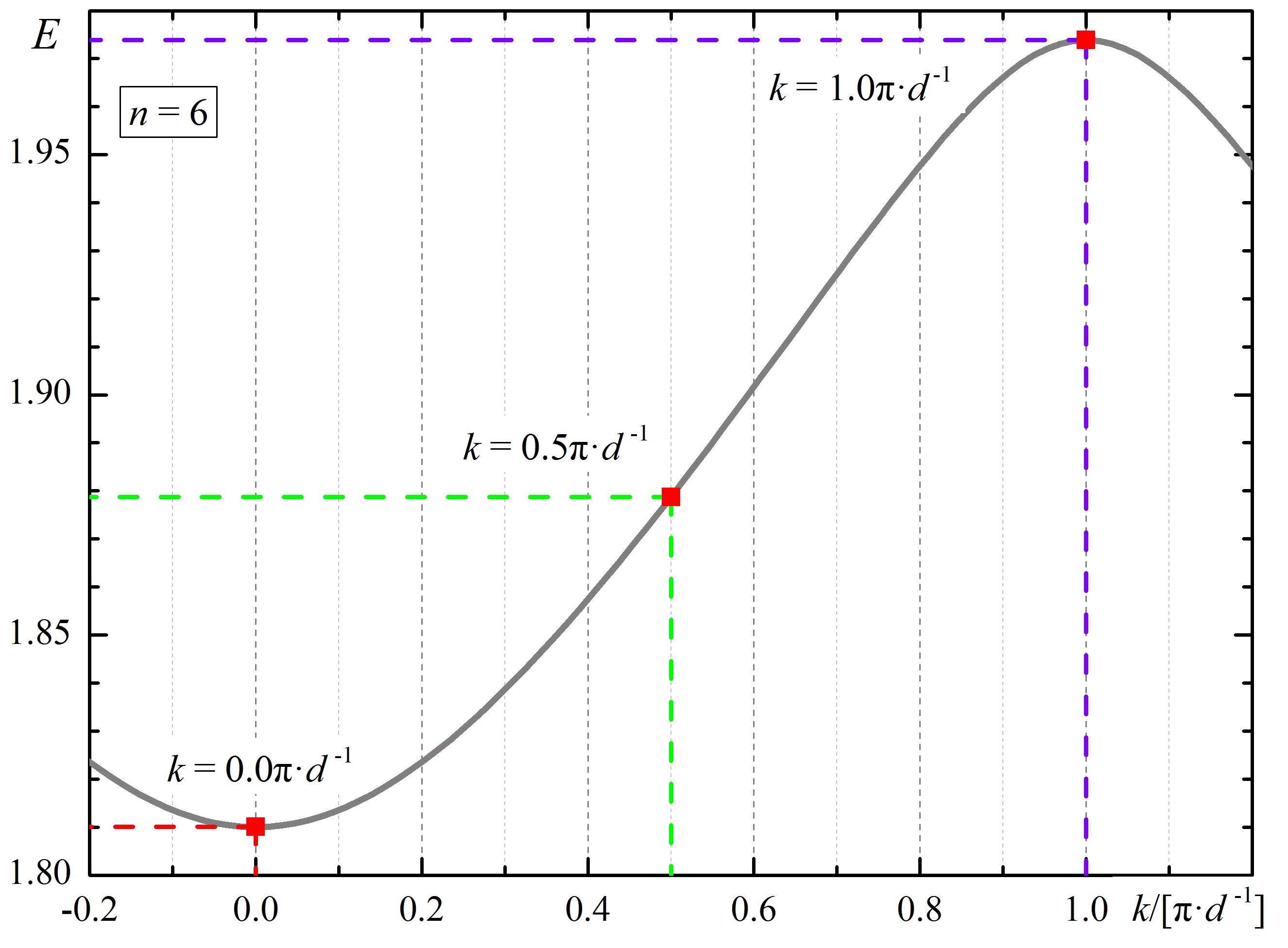}
\par\end{centering}
}

\newpage
\subfloat[(a)]{\begin{centering}
\includegraphics[width=6.2cm]{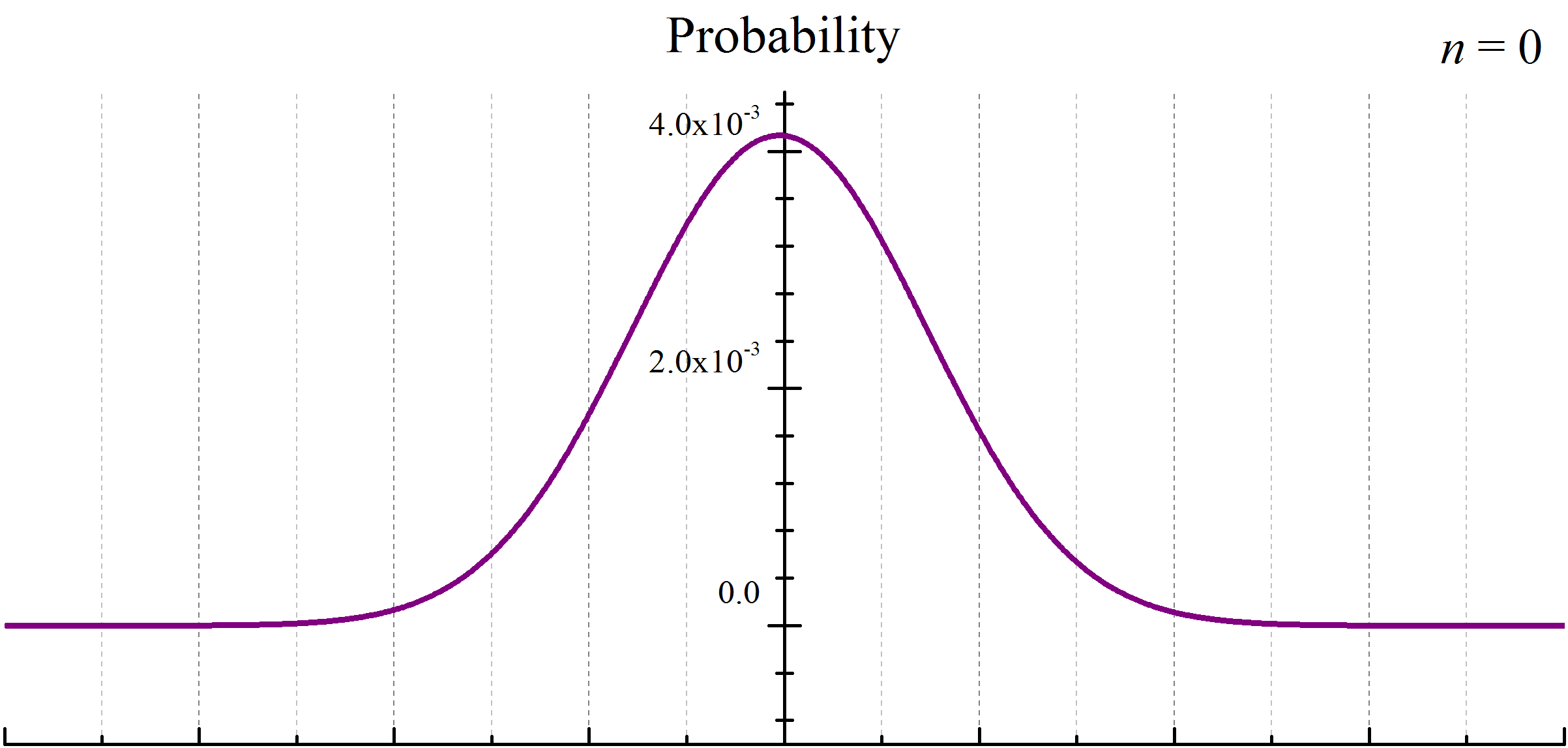}
\par\end{centering}
}

\subfloat[(d)]{\begin{centering}
\includegraphics[width=6.2cm]{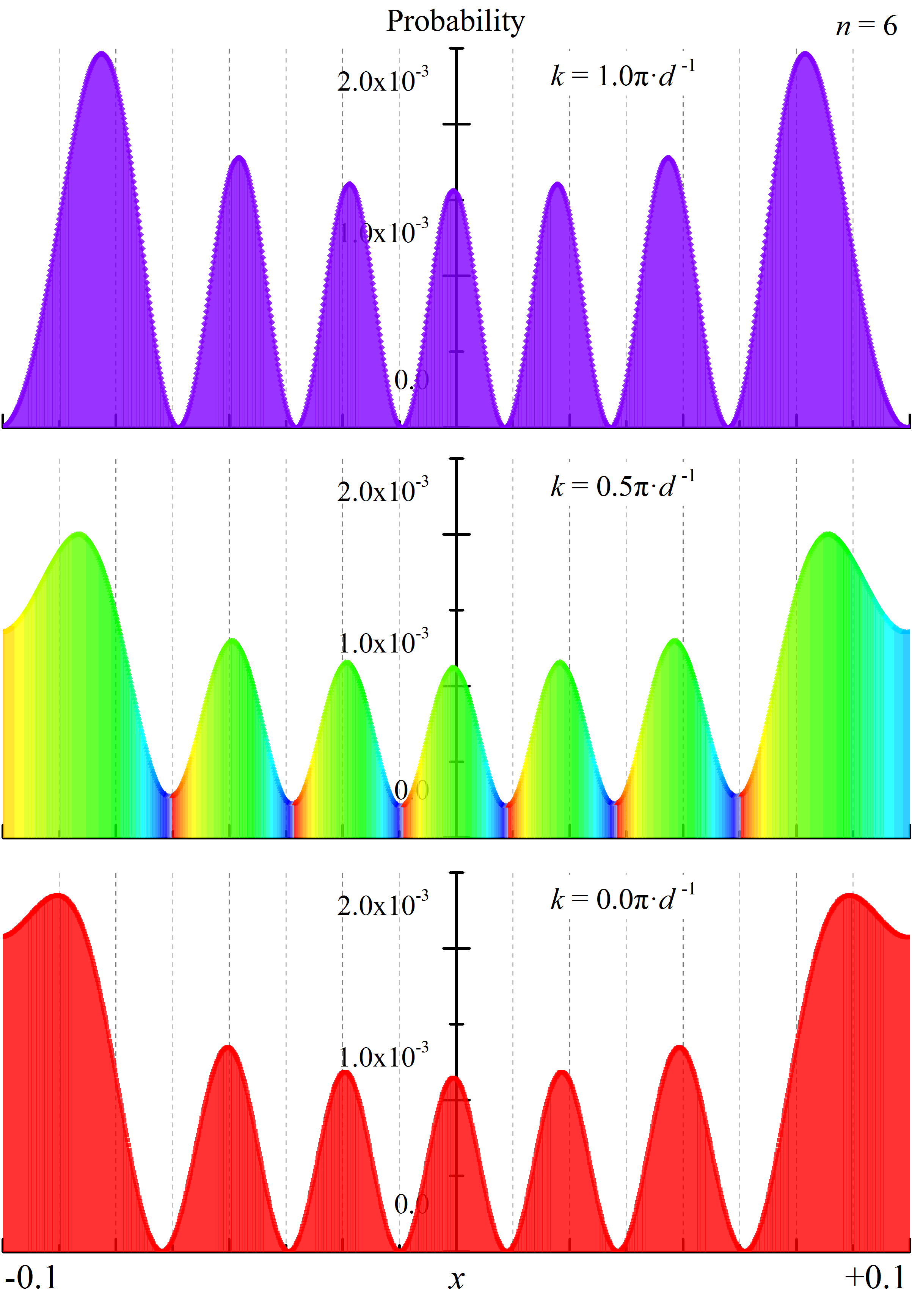}
\par\end{centering}
}
\end{multicols}
\caption{\label{Fig1}(Color online) (a) The probability distribution of stock price in an ordinary harmonic oscillator potential well $V(x)$ when it is in the ground state. (b) The energy levels (blue lines) of an ordinary harmonic potential well $V(x)$ (black solid and blue dashed lines) and the energy bands (red areas) of a spatial-periodic harmonic oscillator potential well $U(x)$ (black solid and red dash-dotted lines). (c) $E$-$k$ relation of the $6-$th conduction band of a spatial-periodic harmonic oscillator potential well $U(x)$. (d) Wave functions of $x$ in the $6-$th conduction band, where the amplitude of curves represents ${{\left| \varphi \left( x \right) \right|}^{2}}$ and the color map represents the phase of $\varphi (x)$.}
\par\end{centering}
\end{figure}

In 1933, a damped harmonic oscillator model \cite{Frisch} was presented. The model presumes that the stock price oscillates and dissipates as a damped harmonic oscillator while being impelled by information that influences the stock market. This model provides an intuitive point of view and introduces physical methods into financial problems. However, according to \cite{PersistentFluctuationsStockMarkets}, it is found that the model cannot explain there existing a persistent small scale of fluctuation of the stock price. Instead, a quantum harmonic oscillator model is introduced \cite{PersistentFluctuationsStockMarkets}. A quantum harmonic model ensures that the volatility $\sigma_x^2(t)$ of a stock is always non-zero, even if there is no information to excite and the oscillator is damped into the ground state. More precisely, the probability distribution ${{\left| \varphi \left( x \right) \right|}^{2}}$ of position $x$ (logarithmic stock price) of the ground state in a harmonic potential well $V(x)=m \omega^2 x^2/2$ is a Gaussian distribution (see Fig. \ref{Fig1}(a) and (b)), that is, ${{\left| \varphi_0 \left( x \right) \right|}^{2}}=\sqrt{m\omega /\pi \hbar }\exp \left( -m\omega {{x}^{2}}/\hbar  \right)$ \cite{QuantumMechanicsCT}, where physical parameters $m,\omega,\hbar$ financially reflects the capital, the oscillating frequency and the uncertainty of irrational transaction of the stock, respectively \cite{QBM}. Besides, we find that the energy of the quantum harmonic oscillator (corresponding to trading volume), i.e., $E=(n+1/2) \hbar \omega$, implies a non-zero ground energy $\hbar \omega/2$. In the ground state, the stock remains standstill, and its price should be determined and equal to its value if all transactions are rational ($\hbar \to 0$), then no rational transactions take place any more ($E \to 0$). Thus the irrational transaction of the stock also leads to a persistent non-zero trading volume.

Since it is known that the daily fluctuation of stock price is limited in $\pm 10\%$ in stock markets of China, a reconsideration of the quantum harmonic oscillator model is required. Before, the limit is considered as a cutoff boundary condition of $x$, that the oscillating particle is limited in an infinite square well potential $-d/2 \le x \le d/2$, for the probability of finding the price out of the width $d$ is absolutely zero. However, this cutoff boundary condition also implies ${{\left| \varphi \left( -d/2 \right) \right|}^{2}}={{\left| \varphi \left( d/2 \right) \right|}^{2}}=0$ to fit the requirement of continuity, thus the probability of reaching the price limit is zero, which seems invalid for stock markets. To modify the boundary condition, ${{\left| \varphi \left( -d/2 \right) \right|}^{2}}={{\left| \varphi \left( d/2 \right) \right|}^{2}}$ is required for the symmetry, but is not necessary to be zero. Such a boundary condition thus indicates $\varphi \left( -d/2 \right)={{e}^{-ikd}}\varphi \left( d/2 \right)$. By making continuation of $\varphi \left( x \right)$ out of the limited width $d$ to infinity and assuming it has a periodic pattern, one has
\begin{equation}
\label{Blochtheorem}
\varphi \left( x \right)={{e}^{-ikd}}\varphi \left( x+d \right),
\end{equation}
which satisfies the one-dimensional Bloch theorem with $k$ a Bloch wave number (phase information of a stock) \cite{SolidStatePhysics}. This periodic boundary condition introduces a spatial-periodic harmonic potential well $U(x)$ (see Fig. \ref{Fig1}(b)). In $-d/2 \le x \le d/2$, we have
\begin{equation}
\label{wavefunction}
\varphi \left( \xi  \right)={{e}^{-{{\xi }^{2}}/2}}\left( A\cdot {{H}_{1}}\left( \xi  \right)+B\cdot {{H}_{2}}\left( \xi  \right) \right),
\end{equation}
where $H_1(\xi)$ and $H_2(\xi)$ are the two independent solutions of the Hermite equation that ${{d}^{2}}H\left( \xi  \right)/d{{\xi }^{2}}-2\xi \cdot dH\left( \xi  \right)/d\xi +\left( 2E/\hbar \omega -1 \right)H\left( \xi  \right)=0$ with $\xi =\beta x=\sqrt{m\omega /\hbar }\cdot x$. From Eq. (\ref{Blochtheorem}), the continuity of $\varphi \left( \xi  \right)$ at $x= \pm d/2$ leads
\begin{equation}
\label{boundarycondition}
\varphi \left( -\frac{\beta d}{2} \right)={{e}^{-ikd}}\varphi \left( \frac{\beta d}{2} \right)\text{,\space\space\space}  \beta {\varphi }'\left( -\frac{\beta d}{2} \right)=\beta {{e}^{-ikd}}{\varphi }'\left( \frac{\beta d}{2} \right).
\end{equation}
Substituting Eq. (\ref{boundarycondition}) into Eq. (\ref{wavefunction}) and noticing coefficients $A$ and $B$ have non-trivial solutions then yield
\begin{eqnarray}
\label{EKrelation}
&&\left[ {{H}_{1}}\left( -\frac{\beta d}{2} \right)-{{e}^{-ikd}}{{H}_{1}}\left( \frac{\beta d}{2} \right) \right]\left[ H_2'\left( -\frac{\beta d}{2} \right)-{{e}^{-ikd}}H_2'\left( \frac{\beta d}{2} \right)+\beta d{{e}^{-ikd}}{{H}_{2}}\left( \frac{\beta d}{2} \right) \right]\nonumber\\
&=&\left[ {{H}_{2}}\left( -\frac{\beta d}{2} \right)-{{e}^{-ikd}}{{H_2}}\left( \frac{\beta d}{2} \right) \right]\left[ H_1'\left( -\frac{\beta d}{2} \right)-{{e}^{-ikd}}H_1'\left( \frac{\beta d}{2} \right)+\beta d{{e}^{-ikd}}{{H}_{1}}\left( \frac{\beta d}{2} \right) \right].
\end{eqnarray}
Within a periodic structure of potential well, the energy levels are broaden as energy bands, and the relation between $E$ and $k$ is already covered in Eq. (\ref{EKrelation}). Based on energy band theory, we can derive analytic solutions of Eq. (\ref{wavefunction}) by using tight binding or free electron approximation, when $E \ll m \omega^2 d^2 /4$ or $E \gg m \omega^2 d^2 /4$, respectively.
\captionsetup[subfigure]{labelformat=empty}

\textit{Tight binding approximation.--}From tight binding approximation, we derive
\begin{equation}
\label{tightbinding}
{{\varphi }_{k}}\left( x \right)=C \sum\nolimits_{s}{{{e}^{s \cdot ikd}}{{\varphi }_{n}}\left( x-sd \right)}\text{,\space\space\space}{{E}_{k}}={{E}_{n}}-\sum\nolimits_{s}{J\left( s \right){{e}^{s \cdot ikd}}},
\end{equation}
where
\begin{equation}
\label{Js}
J\left( s \right)=\int{\varphi _{n}^{*}\left( x-sd \right)\left[ V\left( x \right)-U\left( x \right) \right]}{{\varphi }_{n}}\left( x \right)dx.
\end{equation}
$\{{{\varphi }_{k}}\left( x \right)$, $E_k\}$ and $\{{{\varphi }_{n}}\left( x \right)$, $E_n\}$ of Eqs. (\ref{tightbinding}) and (\ref{Js}) are the wave functions and the energy in $U(x)$ and $V(x)$, related with the Block wave number $k$ and the energy level $n$, respectively. $C$ is an arbitrary normalization factor and $s=0,\pm1,\pm2,\cdots$. From Eq. (\ref{tightbinding}) it is obvious that ${{\varphi }_{k}}\left( x \right)$ is a linear superposition of ${{\varphi }_{n}}\left( x-sd \right)$ with different $s$. When $E \ll m \omega^2 d^2 /4$, ${{\varphi }_{n}}\left( x \right)$ mainly exists in $-d/2 \le x \le d/2$ area and superpositions of ${{\varphi }_{n}}\left( x-sd \right)$ with different $s$ are neglected. Thus we have ${{\varphi }_{k}}\left( x \right) \approx {{\varphi }_{n}}\left( x \right)$ in $-d/2 \le x \le d/2$, but at the edge ($x=\pm d/2$), ${{\varphi }_{k}}\left( x \right)$ is modified by the superpositions of ${{\varphi }_{n}}\left( x-sd \right)$ with different $s$ and behaves differently. To simplify Eq. (\ref{tightbinding}), we only calculate the superpositions of ${{\varphi }_{n}}\left( x \right)$ with itself and its nearby wave functions ${{\varphi }_{n}}\left( x\pm d \right)$. Considering the dependence of the parity of ${{\varphi }_{n}}\left( x \right)$ on $n$, we have $J(+1)=J(-1)$ when $n$ is even and $J(+1)=-J(-1)$ when $n$ is odd. Hence, we derive
\begin{eqnarray}
\label{parity}
{{E}_{k}}&=&{{E}_{n}}-\left| J\left( 0 \right) \right|-2\left| J\left( +1 \right) \right|\cos kd\text{,\space\space\space when }n\text{ is even;}\nonumber\\
{{E}_{k}}&=&{{E}_{n}}-\left| J\left( 0 \right) \right|-2\left| J\left( +1 \right) \right|\sin kd\text{,\space\space\space\space when }n\text{ is odd.}
\end{eqnarray}

\textit{Free electron approximation.--}We assume $U(x)-\overline{U}$ as a perturbation when $E \gg m \omega^2 d^2 /4$, where $\overline{U}=\left\langle  {{\varphi }_{k}} \right|U\left| {{\varphi }_{k}} \right\rangle $ is the average of the potential well. From perturbation theory we have ${{E}_{k}}\approx E_{k}^{(0)}+E_{k}^{(1)}+E_{k}^{(2)},{{\varphi }_{k}}\approx \varphi _{k}^{(0)}+\varphi _{k}^{(1)}$, where
\begin{equation}
\label{E0varphi0}
E_{k}^{(0)}=\frac{\hbar^2 k^2}{2 m}+ \overline{U}\text{,\space\space\space} \varphi _{k}^{(0)}(x)=C  e^{ikx},
\end{equation}
and
\begin{eqnarray}
\label{E2varphi1}
&&E_{k}^{(1)}=0\text{,\space\space\space}\varphi_{k}^{(1)}\left( x \right)=C \sum\nolimits_{s}{K\left( s \right)\frac{2m}{{{\hbar }^{2}}}{{\left[ {{k}^{2}}-{{\left( k+\frac{2\pi s}{d} \right)}^{2}} \right]}^{-1}}{{e}^{i\left(k+2\pi s/d \right)x}}},\nonumber\\
&&E_{k}^{(2)}=\sum\nolimits_{s}{{{\left| K\left( s \right) \right|}^{2}}\frac{2m}{{{\hbar }^{2}}}{{\left[ {{k}^{2}}-{{\left( k+\frac{2\pi s}{d} \right)}^{2}} \right]}^{-1}}},
\end{eqnarray}
where $K\left( s \right)=\left\langle  {{\varphi }_{k+2\pi s/d}^{(0)}} \right|U\left| {{\varphi }_{k}^{(0)}} \right\rangle $ \cite{SolidStatePhysics}. The $0$-th order perturbation in Eq. (\ref{E0varphi0}) implies a behavior of free particle, of which the probability distribution is constant and independent of $x$, while higher order perturbation in Eq. (\ref{E2varphi1}) introduces a modification. It is also implied from Eq. (\ref{E2varphi1}) that when $k=-s\pi/d$ we have $E \to \pm \infty$, which seems invalid. This invalid result is because of the energy degeneracy of the two wave functions with $k=\pm s\pi/d$, respectively. Such a two-fold energy degeneracy will be lifted by perturbation \cite{QuantumMechanicsCT}, which will lead to the formation of forbidden bands.

\section{Influence of price limits on the stock market}
\label{Section3}

To investigate the influence of price limits in more detail, we need to solve Eq. (\ref{wavefunction}) exactly by numerical solution. The energy levels or bands of the potentials are shown in Fig. \ref{Fig1}(b), where $\pm d/2 = \ln(1 \pm 10\%) \approx \pm 10\%$.

When $E \ll m \omega^2 d^2 /4$, the mere difference of energy levels/bands between the ordinary harmonic oscillator potential well and the spatial-periodic one implies that the price has a quasi-harmonic behavior, while when $E \gg m \omega^2 d^2 /4$, the energy space is filled with conduction bands and the price behaves like a free particle, which are just as we derive in Section \ref{Section2}. Besides, the $E$-$k$ relation (see Fig. \ref{Fig1}(c)) of the price in conduction bands will introduce a modification of $\varphi (x)$, as shown in Fig. \ref{Fig1}(d). We find that not only the phase, but also the amplitude of $\varphi (x)$ is modified, which implies that the intra-band probability distribution of the stock price with limits can change observably. As $k$ decreases (and $E$ decreases since $n$ is even), Fig. \ref{Fig1}(d) indicates that it is more probable of finding the stock price at the edge of the limited space ($x=\pm d/2$). The probability distribution when $E$ is relatively small can be explained from a classical view of particle nature, that when a classical particle passes over harmonic periodic barriers, its speed will take minimum value at the edges (the peaks) of the harmonic potential wells, which leads to more probability of finding the particle there. While the probability distribution when $E$ is relatively large can be explained by its wave nature, of which the wave nodes exist at the edges of the wells ($x=\pm d/2$). This intra-band modification of probability distribution is unique, compared with the stock market with no price limits, of which only inter-band influence exists.

\begin{figure}[t]
\begin{centering}

\subfloat[(a)]{\begin{centering}
\includegraphics[width=7.57cm]{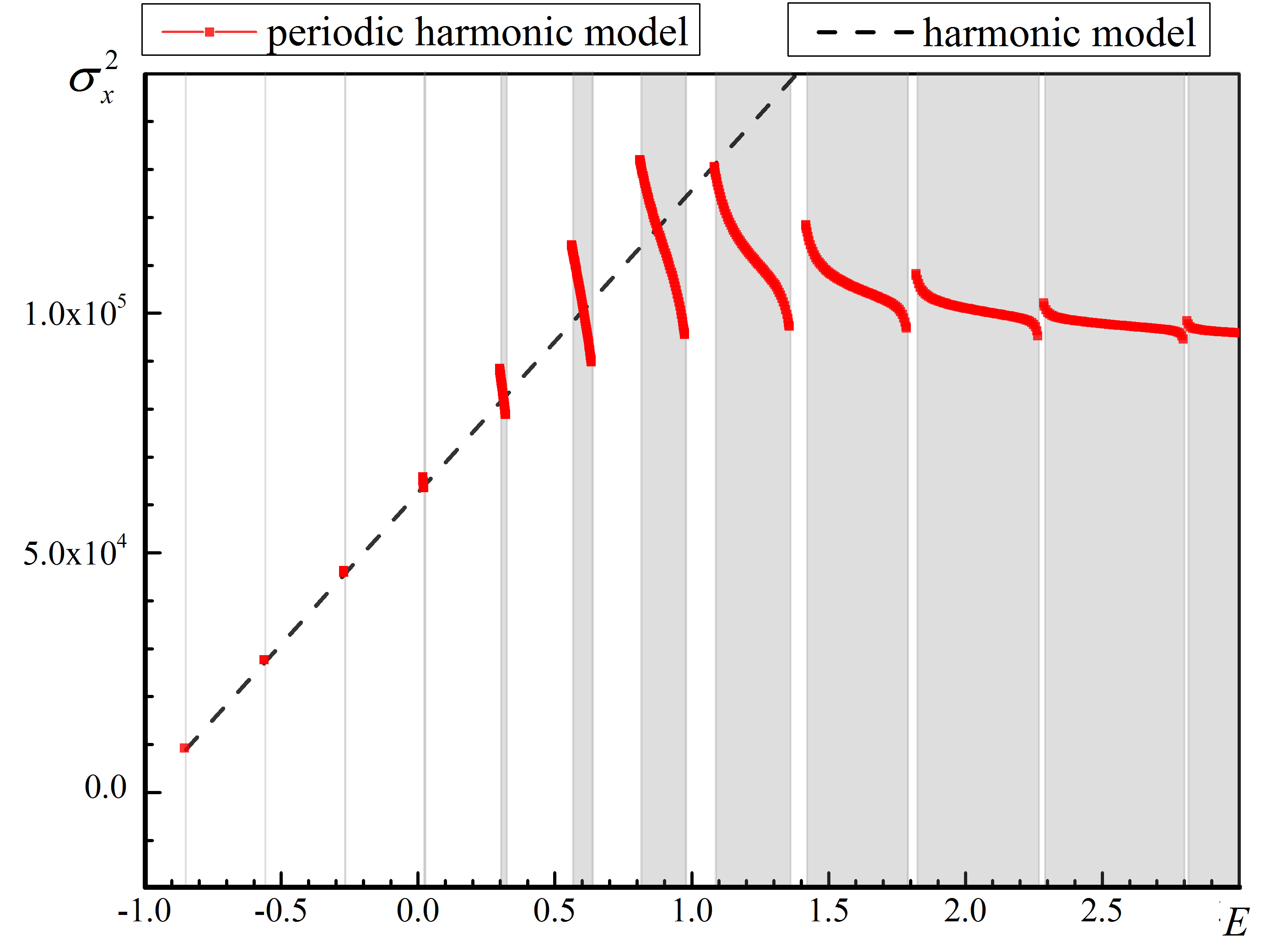}
\par\end{centering}
}\subfloat[(b)]{\begin{centering}
\includegraphics[width=7.6cm]{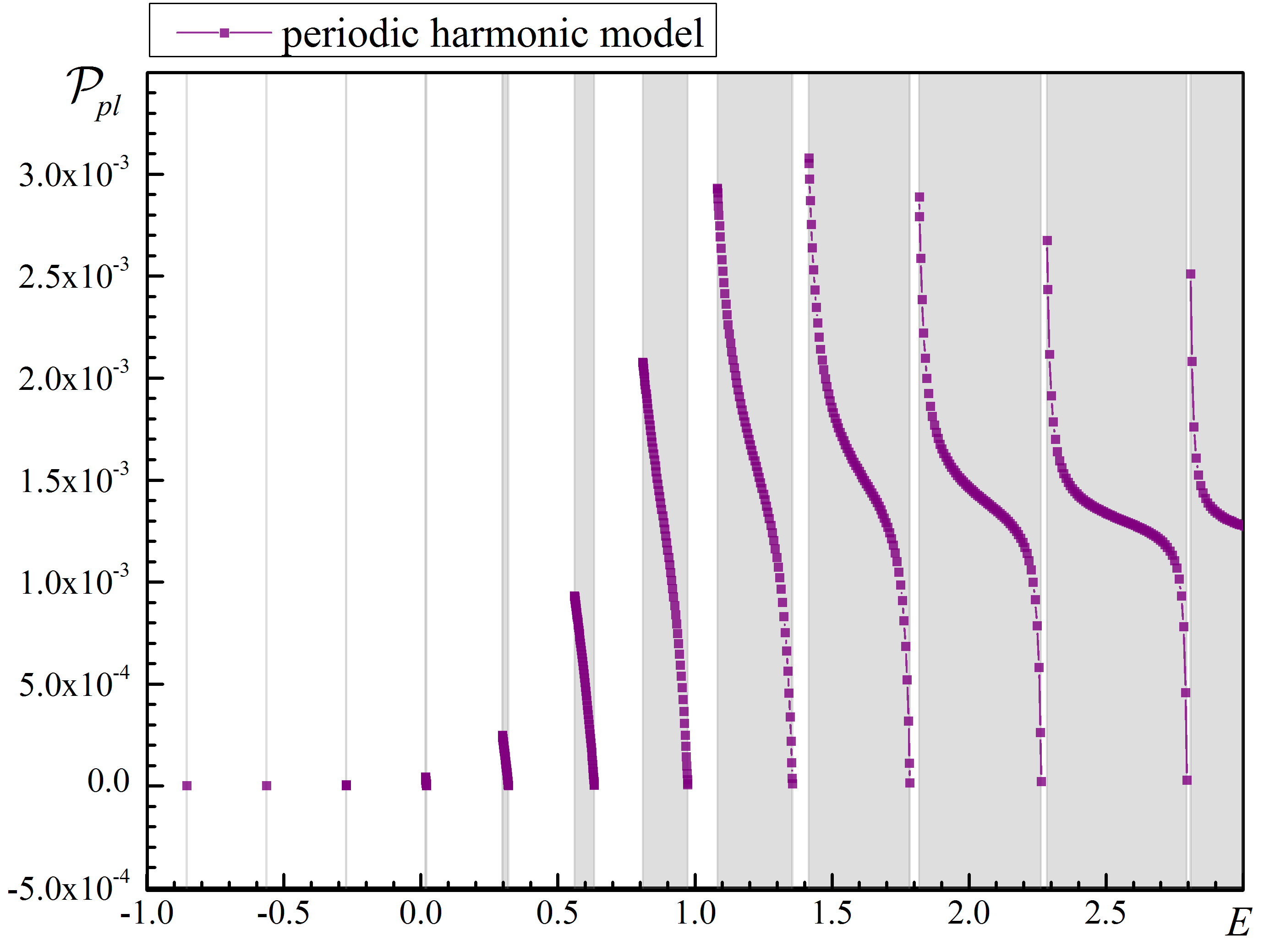}
\par\end{centering}
}

\par\end{centering}
\caption{\label{Fig2}(Color online) (a) The volatility $\sigma_x^2$ as a function of trading volume $E$ of the spatial-periodic harmonic model (red scattered line) is compared with that of the ordinary harmonic model (black dashed line). (b) The probability $\mathcal{P}_{pl}$ of finding the price reaching the price limit (purple scattered line).}
\end{figure}

Furthermore, we investigate the volatility $\sigma_x^2$ of the stock price in details, which is shown in Fig. \ref{Fig2}(a). Compared with the volatility of an ordinary harmonic oscillator that $\sigma_x^2=E/m \omega^2$, the volatility of the spatial-periodic harmonic model is non-linear and much more complicated. When $E \ll m \omega^2 d^2 /4$, $\sigma_x^2$ of the spatial-periodic harmonic model behaves almost the same as the ordinary harmonic model, while when $E \gg m \omega^2 d^2 /4$, $\sigma_x^2$ encounters an inter-band limit and tends to approach a constant. It thus implies it is true that the price limit is able to provide an effective limitation for the volatility of the stock market when the trading volume is high. However, we find that the intra-band behavior of $\sigma_x^2$ is extraordinary, that is, the intra-band $\sigma_x^2$ is negatively correlated with $E$. This unique behavior is indeed related with the $E$-$k$ relation and the intra-band modification of the wave function (see Figs. \ref{Fig1}(c) and \ref{Fig1}(d)). As $E \sim m \omega^2 d^2 /4$, conditions are particularly different, that at the intra-band lower energy edge the volatility $\sigma_x^2$ of the spatial-periodic harmonic model is even greater than that of the ordinary harmonic model. Hence, the price limit cannot provide an effective limitation all the time. Especially when counting the intra-band modification, the price limit can even enhance the volatility of the stock market, which is contrary to perceived purpose. In fact, when a stock price reaches its price limit, e.g., $10\%$, it is hard for stock traders to estimate the real value of the stock correctly, which, for example, may exceeds the limit slightly ($10.1\%$) or heavily ($101\%$). As all possibilities are considered, the average estimation of the value will be generally greater than $10.1\%$, and thus will introduce a greater volatility if the real value happens to exceed slightly ($10.1\%$). This condition happens only when $E \sim m \omega^2 d^2 /4$.

The probability $\mathcal{P}_{pl}$ of reaching the price limit (the edge) is shown in Fig. \ref{Fig2}(b). We see that $\mathcal{P}_{pl}$ as a function of $E$ behaves the same as $\sigma_x^2$ does. However, the intra-band behavior of $\mathcal{P}_{pl}$ is more remarkable that at the intra-band upper energy edge the probability $\mathcal{P}_{pl}$ will always be zero. From a financial trading point of view, a price limitation will enhance the restoring trends of the price. For example, it tends to have more sellers than buyers in the stock market when the price approaches the upper limit $+10\%$, and vice versa. With an intra-band relatively high trading volume, the price is able to be fully limited in the $\pm 10\%$ space. Thus $\mathcal{P}_{pl}$ approaches zero. This condition happens for all energy bands no matter how small or large the trading volume $E$ is.

\section{Quantum phenomena of price limits in the stock market}
\label{Section4}

In this section, we investigate the performance of stocks in the Shanghai Stock Exchange (SSE), for it is one of the largest stock exchanges with price limits set. We mainly use data of $5$-min lines and daily lines of stocks in the SSE, including price volatility and trading volume, and the spatial-periodic harmonic model introduced and studied in Sections \ref{Section2} and \ref{Section3} proves to provide valid explanations for some phenomena in SSE stocks.

\begin{figure}[tbhp]
\begin{multicols}{2}
\begin{centering}
\subfloat[(a)]{\begin{centering}
\includegraphics[width=7.7cm]{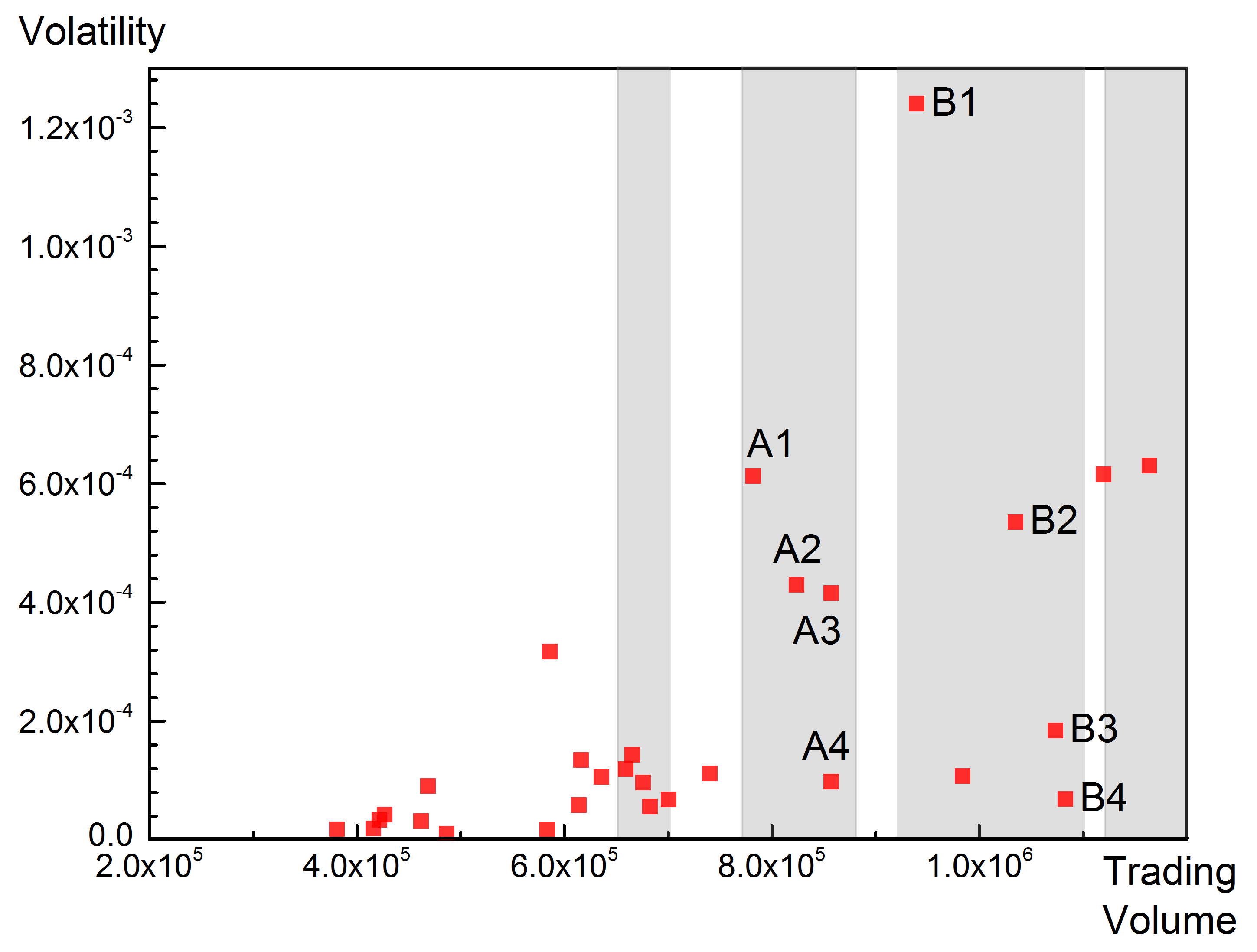}
\par\end{centering}
}

\subfloat[(b)]{\begin{centering}
\includegraphics[width=8.3cm]{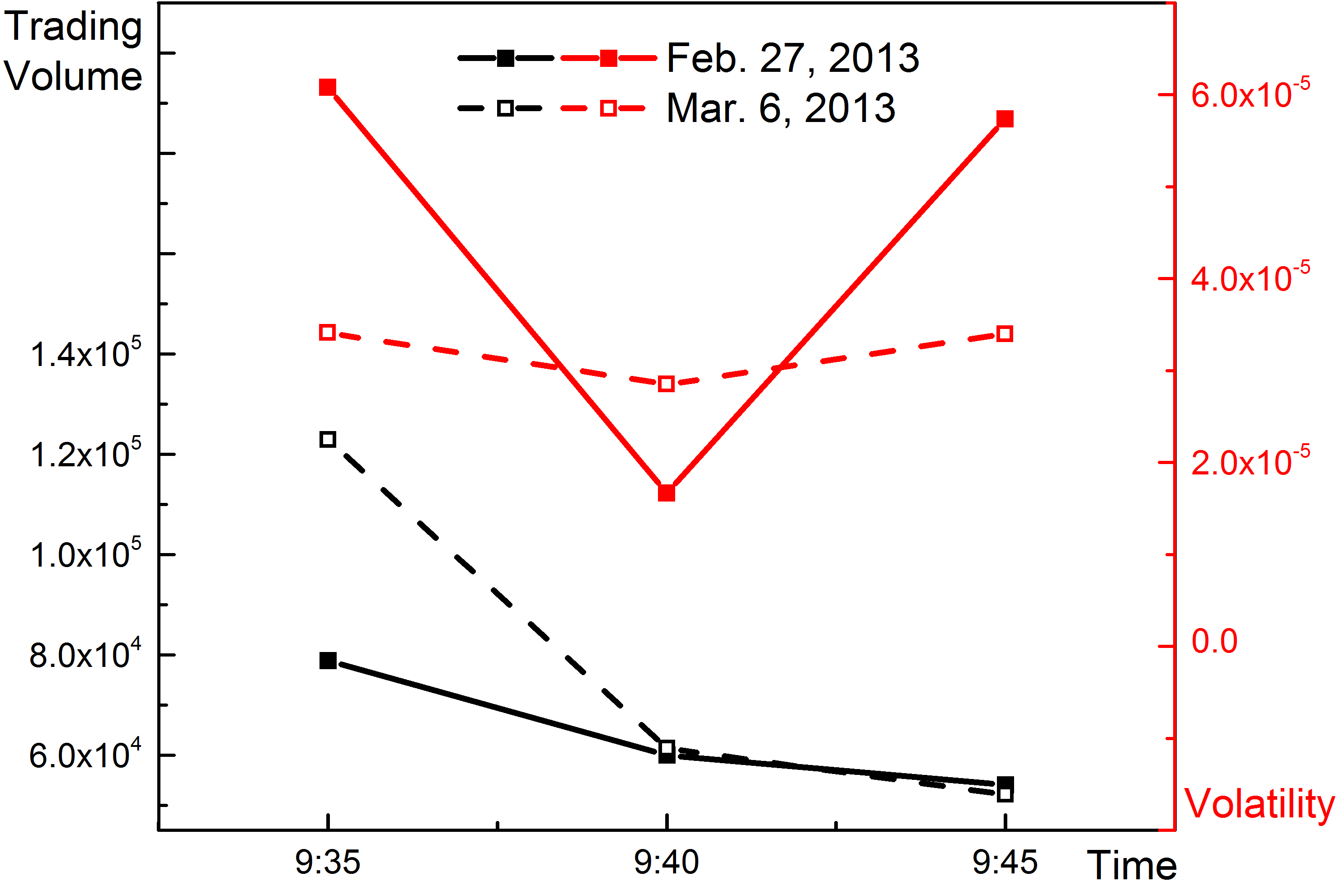}
\par\end{centering}
}

\newpage
\subfloat[(c)]{\begin{centering}
\includegraphics[width=7.7cm]{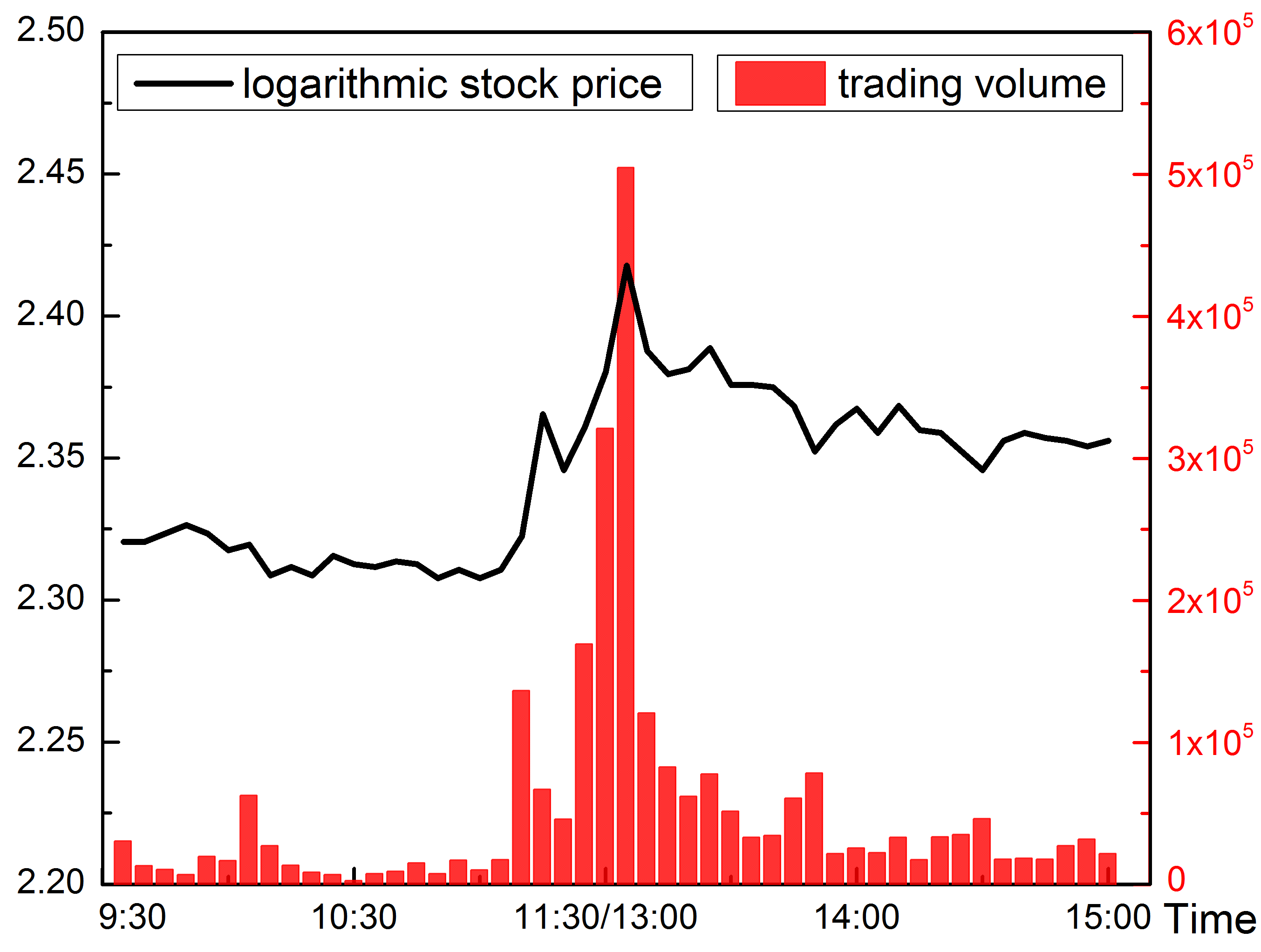}
\par\end{centering}
}
\par\end{centering}
\caption{\label{Fig3}(Color online) (a) The volatility $\sigma_x^2$ as a function of trading volume $E$. The data is abstracted from the daily lines of Ping An Bank Co., Ltd (No. 000001) in the Shanghai Stock Exchange (SSE) from Jan. 14 to Feb. 8, 2013. (b) The volatility $\sigma_x^2$ (red lines) and trading volume $E$ (black lines) abstracted from the $5$-min lines of Ping An Bank Co., Ltd from 09:30 to 09:45, on Feb. 27 (solid lines) and Mar. 6 (dashed lines), 2013. (c) The logarithmic stock price $x$ (black line) and trading volume $E$ (black lines) abstracted from the $5$-min lines of Ping An Bank Co., Ltd on Aug. 16, 2013.}
\end{multicols}
\end{figure}

We first investigate the relation between volatility and trading volume of an arbitrary stock, as shown in Fig. \ref{Fig3}(a). The date of data is chosen specifically, since the stock price of Ping An Bank Co., Ltd once reached $+10\%$ limit on Jan. 14, then oscillated and damped until Feb. 8, after which came the Spring Festival vocation. We find features of energy bands (see Fig. \ref{Fig2}(a)) in Fig. \ref{Fig3}(a), i.e., intra-band negative correlations between $\sigma_x^2$ and $E$. When $E$ is small, the relation between $\sigma_x^2$ and $E$ is not obvious, but it generally shows a positive correlation \cite{PriceChangesandTradingVolume}. When $E$ is large, $\sigma_x^2$ still leads an inter-band increase as a function of $E$. However, the point of the maximum of $\sigma_x^2$, where the stock price reaches $+10\%$ limit, is not related with a maximal $E$. This feature is just shown and studied by us in Fig. \ref{Fig2}(a).

From Fig. \ref{Fig3}(b), we find reverse trends of the stock, i.e., as $E$ decreases, $\sigma_x^2$ will decrease first, but then increase and return to a lower price. This phenomena happen only under conditions that $E$ is not too small and the time period is shorter than about $20$ minutes. We note that this feature can be well explained by damping of the stock price in the energy bands (see Fig. \ref{Fig2}(a)), that is, the former decrease of $\sigma_x^2$ is an inter-band damping, of which the change of $E$ is larger; while the latter increase is an intra-band damping, of which the change of $E$ is smaller. Such an energy band structure exists only when $E \sim m \omega^2 d^2 /4$, and the damping will be re-excited if the time period is too long.

Lastly, we investigate the so-called fat finger incident by China Everbright Securities occurred on Aug. 16, 2013, which caused a large number of stocks reaching $+10\%$ limit that day (see Fig. \ref{Fig3}(c)). Although such an incident was thought to be statistically impossible before, we argue it to be a statistically possible event which is caused by the price limits of stock exchanges of China. It is implied from Figs. \ref{Fig2}(a) and \ref{Fig2}(b) that when $E \sim m \omega^2 d^2 /4$, an inter-band excitation for the stock price to jump from the top of the lower conduction band to the bottom of the upper one will increase $\sigma_x^2$ and $\mathcal{P}_{pl}$ considerably, and the fluctuation of price becomes stronger. Hence, the fat finger event should not be treated as an improbable incidents but a reflection of the stronger fluctuation, which warns us to propose relevant rules and policies to prevent such events.

We note the explanations of stock phenomena shown in Fig. \ref{Fig3} are directly derived from the quantum spatial-periodic harmonic model, and are generally related with the uncertainty $\hbar$ of irrational transactions. If $\hbar \to 0$, the spatial-periodic harmonic model degenerates to a classical one and implies a continuous $E$ with no energy bands contained. Hence, we say such studied phenomena are in fact quantum phenomena of the stock market, i.e., irrational phenomena \cite{QBM}.

\section{Discussion and Conclusions}
\label{Section5}

In before sections, we introduce the energy $E$ as a parameter reflecting the trading volume of stocks, and it proves to be a valid presumption due to our study of the stock data. However, the exact scaling of trading volume needs a more careful consideration for it is not simple to derive an exact relation between $E$ and trading volume. Nonetheless, we suggest that trading volume may be linearly related with the number of states $n$. By considering the density of states, it is suggested that when $E \ll m \omega^2 d^2 /4$, a harmonic oscillator implies $E \propto n$; when $E \gg m \omega^2 d^2 /4$, a free particle implies $E \propto n^2$. Since the positive correlation between $E$ and trading volume is always ensured, our results still stay valid.

In conclusion, we investigate the behavior of stocks in a price-limited stock market by purposing a quantum spatial-periodic harmonic model in this work. The stock price is presumed oscillating and damping in a spatial-periodic harmonic oscillator potential well. We introduce a theoretical model and study the non-linear relevant features of volatility $\sigma_x^2$, trading volume $E$, etc. When the trading volume is small, the stock price behaves like a harmonic oscillator, while a large trading volume implies a free particle behavior. Besides, the structure of energy bands is found in the spatial-periodic harmonic model, from which it is indicated that the price limit is able to provide an effective limitation of volatility when the trading volume is large enough, but it will increase the volatility on the contrary if within a certain regime of the trading volume. Furthermore, the numerical solution of the energy bands implies that $\sigma_x^2$ and $E$ of stocks has not only a general inter-band positive correlation but also an intra-band negative correlation. In addition, the probability $\mathcal{P}_{pl}$ of reaching the price limit always approaches zero for specific $E$ (at the upper edges of energy bands). We further investigate the performance of stocks in Shanghai Stock Exchange of China where price limits are set. Some irrational phenomena, noted by us as quantum phenomena, are found in the stock markets, e.g., negative correlations between $\sigma_x^2$ and $E$, reverse trends of $\sigma_x^2$ and its abnormal increase. These phenomena are well explained by the different inter-band and intra-band features and damping (exciting) transitions between them. We remark that the quantum spatial-periodic harmonic model is practicable due to its understandable physical characteristics, and it proves to be a practicable model for price-limited stock markets.



%% References


\bibliographystyle{model1-num-names}
\bibliography{QPHM}






% \end{thebibliography}

\end{document}